\documentclass[twocolumn,pra,showpacs,superscriptaddress,amssymb,amsmath,amsmath]{revtex4-1}
\usepackage{graphicx}
\usepackage{epstopdf}
\usepackage{bm}
\usepackage{hyperref}

\hypersetup{%
   pdfpagemode=None, %FullScreen,n
   pdfstartpage=1,
   pdfmenubar=true,
   pdftoolbar=true,
   colorlinks = true,
   linkcolor=blue,
   citecolor=blue,
   bookmarksopen=false
 }

\newcommand{\be}{\begin{equation}}
\newcommand{\ee}{\end{equation}}

\begin{document}

\title{Prospects for ultracold polar and magnetic chromium--closed-shell-atom molecules}

\author{Micha\l~Tomza}
\email{michal.tomza@chem.uw.edu.pl}
\affiliation{Faculty of Chemistry, University of Warsaw, Pasteura 1, 02-093 Warsaw, Poland}

\date{\today}

\begin{abstract}

The properties of the electronic ground state of the polar and paramagnetic chromium--closed-shell-atom molecules have been investigated. State-of-the-art \textit{ab initio} techniques have been applied to compute the potential energy curves for the chromium--alkaline-earth-metal-atom, CrX (X = Be, Mg, Ca, Sr, Ba), and chromium--ytterbium, CrYb, molecules in the Born-Oppenheimer approximation for the high-spin $X^7\Sigma^+$ electronic ground state. The spin restricted open-shell coupled cluster method restricted to single, double, and noniterative triple excitations, RCCSD(T), was employed and the scalar relativistic effects within the Douglas-Kroll-Hess Hamiltonian or energy-consistent pseudopotentials were included. The permanent electric dipole moments and static electric dipole polarizabilities were computed. The leading long-range coefficients describing the dispersion interaction between the atoms at large interatomic distances, $C_6$, are also reported.
Molecules under investigation are examples of species possessing both large magnetic and electric dipole moments making them potentially interesting candidates for ultracold many-body physics studies.

\end{abstract}


\maketitle

\section{Introduction}

The research on atoms and molecules at ultralow temperatures addresses the most fundamental questions of quantum mechanics~\cite{ColdAtomsandMolecules}.  
The field of ultracold matter started with gases of alkali-metal atoms and for many years has been restricted to these species~\cite{JulienneRMP99}. All ultracold ($T<1$mK) molecules in the absolute rovibrational ground state, produced to this day, consist of alkali-metal atoms~\cite{WeidemullerCR12}. Nevertheless, recent success in cooling and Bose-Einstein condensating the highly magnetic $^{52}$Cr~\cite{PfauPRL05}, $^{168}$Er~\cite{FerlainoPRL12}, and $^{164}$Dy~\cite{LevPRL11} atoms or closed-shell ${}^{40}$Ca~\cite{KraftPRL09}, ${}^{84}$Sr~\cite{StellmerPRL09,EscobarPRL09}, ${}^{86}$Sr~\cite{StellmerPRA10}, ${}^{88}$Sr~\cite{MickelsonPRA10}, ${}^{170}$Yb~\cite{FukuharaPRA07}, and ${}^{174}$Yb~\cite{TakasuPRL03} atoms allow to consider them as candidates for forming ultracold molecules.

Heteronuclear molecules possessing permanent electric dipole moment are promising candidates for numerous applications including quantum computing, quantum simulations, many-body physics, ultracold controlled chemistry, precision measurements, and tests of fundamental laws \cite{ColdMolecules}.
Heteronuclear molecules formed from atoms with large magnetic dipole moments could possess both magnetic and electric dipole moments that would provide an additional knob to control the quantum dynamics with both magnetic and electric fields~\cite{QuemenerCR12}.

Recently there has been an increased interest in the study of ultracold mixtures of open-shell and closed-shell atoms. Ultracold mixtures of Li and Yb~\cite{DoylePRL11,IvanovPRL11}, Rb and Yb~\cite{BaumerPRA11,NemitzPRA09}, Cs and Yb~\cite{Cornish13}, and Rb and Sr~\cite{Schreck13} atoms have been investigated experimentally. Open-shell Li--alkali-earth-metal-atom~\cite{KotochigovaJCP11,HadaPRA11}, LiYb~\cite{ZhangJCP10,GopakumarJCP10,BruePRL12}, alkali-metal-atom--Sr~\cite{GueroutPRA10} and RbSr~\cite{ZuchowskiPRL10} molecules have been explored theoretically.
Although the properties of the alkali-metal-atom--closed-shell-atom molecules could be tuned with external electric and magnetic fields e.g.~by controlling the spin-dependent long-range interactions, the intermolecular magnetic dipole-dipole interaction resulting from their magnetic dipole moments is too small to compete against the electric dipole-dipole interaction or short-range chemical forces and to influence the many-body dynamics. To explore the impact of the intermolecular magnetic dipole-dipole interaction on the properties of ultracold molecular gas, molecules formed from the highly magnetic atoms such as Cr(${}^7S$), Eu(${}^8S$), Er(${}^3H$), or Dy(${}^5I$) should be considered.   

A high-spin spherically symmetric $S$-state chromium atom is a natural candidate for the formation of a molecule possessing a large magnetic dipole moment.
Properties of the electronic ground state of the chromium--alkali-metal-atom molecules have been investigated theoretically~\cite{SadeghpourPRA10, JeungJPB10} and the CrRb molecule was proposed as a candidate for a molecule with both large magnetic and electric dipole moments~\cite{SadeghpourPRA10}.  
The two-species magneto-optical trap (MOT) for the Cr and Rb atoms was realized in 2004~\cite{PfauJMO04} but the operation of superimposed MOTs was limited by the photoionization of the excited state of the Rb atoms by the Cr cooling-laser light. Unfortunately, the same trap losses are expected for the mixtures of chromium with other alkali-metal atoms. Since the ionization potentials of alkali-earth-metal atoms are at least by 10,000$\,$cm$^{-1}$ larger than for alkali-metal atoms of similar size, this problem will not occur for the two-species MOT with chromium and alkali-earth-metal atoms or alkali-earth-metal-like Yb atoms. 

There are advantages of using a closed-shell ${}^1S$ atom as a partner of ${}^7S$ chromium atom for the formation of a highly magnetic open-shell molecule. First of all the resulting electronic structure of such a system is relatively simple. There is only one electronic state dissociating into ground-state closed-shell and ground-state chromium atoms. The zero internal orbital angular momentum of both atoms implies the $\Sigma$ symmetry of the electronic ground state.
Therefore, there is no anisotropy of the interaction between the atoms that could lead to the fast Zeeman relaxation and losses in the formation process of the magnetic molecules from highly magnetic atoms with large orbital angular momentum~\cite{KotochigovaPRL12}.
Finally, the molecule inherits the large magnetic dipole moment of the chromium atom, $d_m=6\mu_B$.

Until recently, the most efficient method of forming ultracold molecules, that is magnetoassociation within the vicinity of the Feshbach resonances followed by the stimulated Raman adiabatic passage (STIRAP), was believed to be restricted to alkali-metal-atom dimers~\cite{NiScience08,DeiglmayrPRL08}. However, recent works by \.Zuchowski et al.~\cite{ZuchowskiPRL10} and Brue and Hutson~\cite{BruePRL12} suggest that it is possible to form open-shell-atom--closed-shell-atom molecules by magnetoassociation using the interaction-induced variation of the hyperfine coupling constant.

For the above reasons, in the present work we investigate the properties of the electronic ground state of the chromium--alkaline-earth-metal-atom and chromium--ytterbium molecules. 
To the best of our knowledge, the chromium--closed-shell-atom molecules have not yet been considered theoretically or experimentally, except the recent work on the Feshbach resonances in the Cr and Yb atoms mixture by \.Zuchowski~\cite{ZuchowskiCrYb} and buffer gas cooling of the Cr atoms with a cryogenically cooled helium~\cite{WeinsteinPRA98}. Here we fill this gap and report the \textit{ab initio} properties of the $^7\Sigma^+$ electronic ground states of the chromium--alkaline-earth-metal-atom and chromium--ytterbium molecules paving the way towards a more elaborate study of the formation and application of these polar and magnetic molecules.

The plan of our paper is as follows. Section~\ref{sec:theory} describes the theoretical methods used in the \textit{ab initio} calculations. Section~\ref{sec:results} discusses the potential energy curves and properties of the chromium--alkali-earth-metal-atom and chromium--ytterbium molecules in the rovibrational ground state and analyzes the completeness and accuracy of the applied \textit{ab initio} methods.
It also surveys the characteristic length scales related to the intermolecular magnetic and electric dipolar interactions.
Section~\ref{sec:summary} summarizes our paper.

\section{Computational Details}
\label{sec:theory}

The chromium--closed-shell-atom molecules are of open-shell nature, therefore we have calculated the potential energy curves in the Born-Oppenheimer approximation using the spin restricted open-shell coupled cluster method restricted to single, double, and noniterative triple excitations, starting from the restricted open-shell Hartree-Fock (ROHF) orbitals, RCCSD(T)~\cite{KnowlesJCP99}. The interaction energies have been obtained with the supermolecule method correcting the basis-set superposition error~\cite{BSSE}:
\begin{equation}
V_\textrm{CrX}=E_\textrm{CrX}-E_\textrm{Cr}-E_\textrm{X}\,,
\end{equation}
where $E_\textrm{CrX}$ denotes the energy of the dimer, and $E_\textrm{Cr}$ and $E_\textrm{X}$ are the energies of the monomers computed in the dimer basis. 

The scalar relativistic effects in the calculations for the CrBe, CrMg, and CrCa molecules were included by employing the second order Douglas-Kroll-Hess, DKH, Hamiltonian~\cite{ReiherTCA06}, whereas for the CrSr, CrBa, and CrYb molecules the relativistic effects were accounted for by using small-core fully relativistic energy-consistent pseudopotentials, ECP, to replace the inner-shells electrons~\cite{DolgCR12}. We used the pseudopotentials to introduce the relativistic effects for heavier molecules instead of using the Douglas-Kroll-Hess Hamiltonian, because it allowed to use larger basis sets to describe valence electrons and modeled the inner-shells electrons density as accurately as the high quality atomic calculation used to fit the pseudopotentials.

In all calculations for the CrBe, CrMg, and CrCa molecules the augmented correlation consistent polarized Valence Quintuple-$\zeta$ quality basis sets, aug-cc-pV5Z, were used. The Be and Cr atoms were described with the aug-cc-pV5Z-DK basis sets~\cite{BalabanovJCP05}, whereas for the Mg and Ca atoms, the cc-pV5Z-DK and cc-pV5Z basis sets~\cite{KoputJPCA}, respectively, were augmented at first.
In all calculations for CrSr, CrBa, and CrYb the pseudopotentials from the Stuttgart library were employed. The Cr atom was described by the ECP10MDF pseudopotential~\cite{DolgJCP87} and the $[14s13p10d5f4g3h]$ basis set with coefficients taken from the aug-cc-pVQZ-DK basis~\cite{BalabanovJCP05}.
The Sr atom was described with the ECP28MDF pseudopotential~\cite{LimJCP06} and the $[14s11p6d5f4g]$ basis set obtained by augmenting the basis set suggested in Ref.~\cite{LimJCP06}.
The Ba atom was described with the ECP46MDF pseudopotential~\cite{LimJCP06} and the $[13s12p6d5f4g]$ basis set obtained by augmenting the basis set suggested in Ref.~\cite{LimJCP06}.
The Yb atom was described with the ECP28MDF pseudopotential~\cite{DolgTCA98} and the $[15s14p12d11f8g]$ basis set~\cite{DolgTCA98}.
In all calculations the basis sets were augmented by the set of $[3s3p2d1f1g]$ bond functions~\cite{midbond}.

The permanent electric dipole moments:
\begin{equation}
d_i=\langle\Psi_{\textrm{CrX}}| \hat{d}_i|\Psi_{\textrm{CrX}}\rangle=\left.\frac{\partial E_{\textrm{CrX}}(F_i)}{\partial F_i}\right|_{F_i=0}\,,
\end{equation}
where $\hat{d}_i$, $i=x,y$ or $z$, denotes the $i$th component of the electric dipole moment operator and static electric dipole polarizabilities:
\begin{equation}
\alpha_{ij}=\left.\frac{\partial^2 E_{\textrm{CrX}}(\vec{F})}{\partial F_{i}\partial F_{j}}\right|_{\vec{F}=0}\,,
\end{equation}
were calculated with the finite field method. 
The dipole moments and the polarizabilities were obtained with three-point and five-point approximations of the first and second derivatives, respectively. The $z$ axis was chosen along the internuclear axis and oriented from the closed-shell to the chromium atom.
 
The interaction potential between two neutral atoms in the electronic ground state is asymptotically given by the dispersion interaction of the form~\cite{HeijmenMP96}:
\begin{equation}\label{eq:E6}
V_{\textrm{CrX}}(R)=-\frac{C_6}{R^6}+\dots\,,
\end{equation}
where the leading $C_6$ coefficient given by:
\begin{equation}\label{eq:C6}
C_6=\frac{3}{\pi}\int_0^\infty \alpha_{\textrm{Cr}}(i\omega)\alpha_\textrm{X}(i\omega)d\omega\,,
\end{equation}
is the integral over the dynamic polarizabilities of the Cr and X atoms at an imaginary frequency, $\alpha_{\textrm{Cr/X}}(i\omega)$. The dynamic electric dipole polarizability is given by:
\begin{equation}\label{eq:alpha}
\alpha_{\textrm{X}}(\omega)=\sum_n \frac{f^{\textrm{X}}_{0n}}{\omega^2_{\textrm{X},0n}-\omega^2}\,,
\end{equation}
where  $f^{\textrm{X}}_{0n}$ denotes the oscillator strength between the atomic ground
state and the $n$th atomic excited state, and $\omega_{\textrm{X},n0}$ is the excitation energy to that state.

The dynamic electric dipole polarizabilities at an imaginary frequency of the alkali-earth-metal atoms were taken from the work by Derevianko et al.~\cite{DerevienkoADNDT10}, whereas the dynamic polarizability of the ytterbium atom was obtained by using the explicitly connected representation of the expectation value
and polarization propagator within the coupled cluster method~\cite{MoszynskiCCCC05} and the best approximation XCCSD4 proposed by Korona and collaborators~\cite{KoronaMP06}.
The dynamic polarizability of the chromium atom was constructed as a sum over states, Eq.~\ref{eq:alpha}. The oscillator strengths and energy levels for the discrete transitions were taken from the NIST Atomic Spectra Database~\cite{NIST}, whereas the contribution form the bound-continuum transitions were included as a sum over oscillator strengths to quasi-bound states obtained within the MRCI method.

All calculations were performed with the MOLPRO package of \textit{ab initio} programs~\cite{Molpro}.

\section{Numerical results and discussion}
\label{sec:results}

\begin{table*}[]
\caption{Spectroscopic characteristics: equilibrium bond length, $R_e$, well depth, $D_e$, harmonic frequencies, $\omega_0$, number of bound vibrational states, $N_\upsilon$, and long-range dispersion coefficient, $C_6$, of the $X^7\Sigma^+$ ground electronic state and rotational constant $B_0$, electric dipole moment, $d_0$, average polarizability, $\bar{\alpha}_0$, and polarizability anisotropy, $\Delta\alpha_0$, for the rovibrational ground level of the $X^7\Sigma^+$ ground electronic state of the CrBe, CrMg, CrCa, CrSr, CrBa, and CrYb molecules. $\tilde{C}_6$ is the coefficient for the intermolecular dispersion interaction between molecules in the ground rovibrational level.}
\begin{ruledtabular}
\begin{tabular}{lrrrrrrrrrr}
Molecule & $R_e\,$(bohr) & $D_e\,$(cm$^{-1}$) &  $\omega_0\,$(cm$^{-1}$) & $N_\upsilon$ & $B_0\,$(cm$^{-1}$)&  $d_0\,$(D) & $\bar{\alpha}_0\,$(a.u.) & $\Delta\alpha_0\,$(a.u.) & $C_6\,$(a.u.)&  $\tilde{C}_6\,$(a.u.) \\
\hline
$^{52}$Cr$^{9}$Be   & 4.56 & 4018 & 319  & 29 & 0.377 &  1.43 & 121.4 & 102.3 & 383 & $1.5\cdot 10^4$\\
$^{52}$Cr$^{24}$Mg  & 5.50 & 2441 & 141  & 39 & 0.121 &  0.10 & 170.8 & 158.3 & 667 & $1.1\cdot 10^4$\\
$^{52}$Cr$^{40}$Ca  & 5.94 & 3548 & 136  & 62 & 0.076 & -0.76 & 248.9 & 178.1 & 1232 & $2.7\cdot 10^4$\\
$^{52}$Cr$^{88}$Sr  & 6.15 & 3649 & 107  & 75 & 0.049 & -1.48 & 283.5 & 176.1 & 1488 & $1.2\cdot 10^5$\\
$^{52}$Cr$^{138}$Ba & 6.22 & 4776 & 106  & 94 & 0.041 & -2.67 & 345.6 & 121.9 & 1905 & $1.1\cdot 10^6$ \\
$^{52}$Cr$^{174}$Yb & 6.05 & 2866 & 87.8 & 73 & 0.041 & -1.19 & 242.9 & 178.9 & 1195 & $6.5\cdot 10^4$\\
\end{tabular}
\label{tab:Spectro}
\end{ruledtabular}
\end{table*}

\subsection{Potential energy curves}

\begin{figure}[t!]
\begin{center}
\includegraphics[width=\columnwidth]{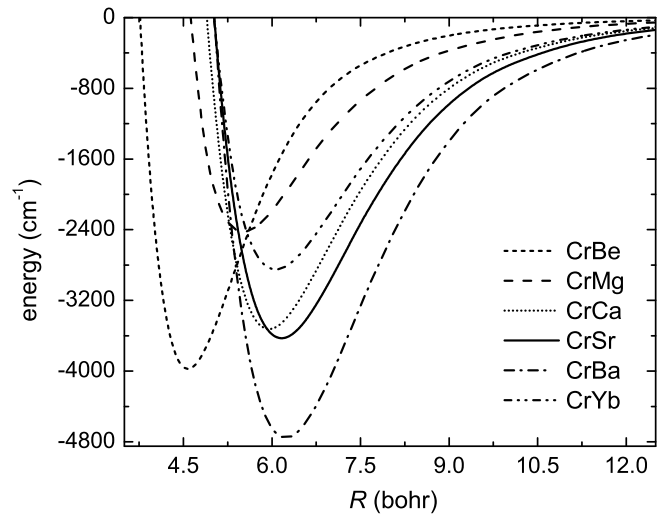}
\end{center}
\caption{Potential energy curves of the $X^7\Sigma^+$ electronic ground state of the CrBe, CrMg, CrCa, CrSr, CrBa, and CrYb molecules.}
\label{fig:Curves}
\end{figure}

The computed potential energy curves of the $X^7\Sigma^+$ electronic ground state of the CrBe, CrMg, CrCa, CrSr, CrBa, and CrYb molecules are presented in Fig.~\ref{fig:Curves} and the corresponding long-range $C_6$ coefficients are reported in Table~\ref{tab:Spectro}.
The equilibrium distances, $R_e$, and well depths, $D_e$, are also collected in Table~\ref{tab:Spectro}. 

An inspection of Fig.~\ref{fig:Curves} reveals that all potential energy curves show a smooth behavior with well defined minima.
The well depths of the chromium--alkaline-earth-metal-atom and chromium--ytterbium molecules are significantly larger (by a factor of two to four) than those of the Van der Waals type homonuclear alkaline-earth-metal-atom~\cite{AmitayPRL11,SkomorowskiJCP12,BusseryPRA03} or ytterbium molecules~\cite{BuchachenkoEPJD07}. The largest dissociation energy is 4723$\,$cm$^{-1}$ for the CrBa molecule and the smallest one is 2371$\,$cm$^{-1}$ for the CrMg molecule. The equilibrium distances take values between 4.56$\,$bohr for the CrBe molecule up to 6.22$\,$bohr for the CrBa molecule.
The dissociation energies and equilibrium distances of the investigated molecules are systematically increasing with the increasing mass of the alkaline-earth-metal atom, except for the dissociation energy of the CrBe molecule which is much larger than expected. However, the much stronger binding energy and shorter equilibrium distance of the CrBe molecule is not surprising when we know that the beryllium dimer has an unexpectedly strong bonding interaction, substantially stronger and shorter than those between other similarly sized closed-shell atoms \cite{PatkowskiScience09}.  
The $C_6$ coefficients are rather small and typical for the Van der Waals type molecules.

The existence of the potential energy crossing between the $X^7\Sigma^+$ state and some lower spin state is very unlikely. 
The lower spin states are higher in energy because either they are connected with the excited states of chromium and then the interaction energy is of the same order as for the ground state or they are connected with the excited states of the closed-shell atom with excitation energies much larger than the depth of potentially deep potential energy curves.    
Therefore, the ultracold collisions between the ground-state chromium and close-shell atoms should fully be described on the $X^7\Sigma^+$ potential energy curve.

\textit{Ab initio} potentials were used to calculate the rovibrational spectra of the $X^7\Sigma^+$ electronic ground states for the molecules consisting of the most abundant isotopes. The harmonic frequencies, $\omega_0$, and the numbers of the supported bound states for the angular momentum $J=0$, $N_\upsilon$,  are reported in Table~\ref{tab:Spectro}. Rotational constants for the rovibrational ground state, $v=0,J=0$, were also calculated and are reported in Table~\ref{tab:Spectro}.

\begin{figure}[t!]
\begin{center}
\includegraphics[width=\columnwidth]{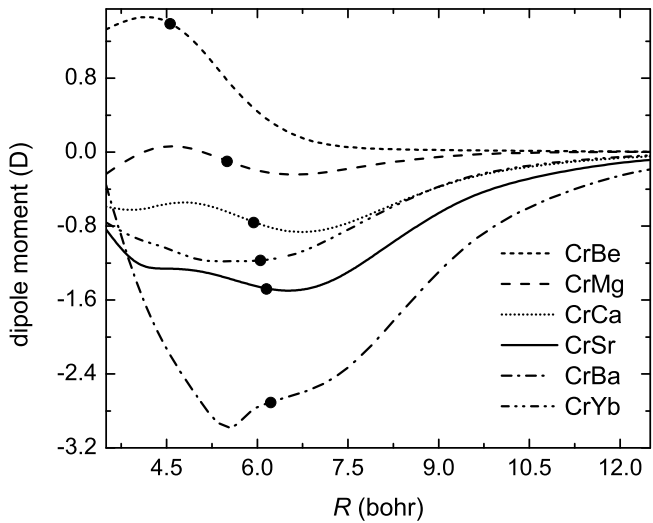}
\end{center}
\caption{Permanent electric dipole moments of the $X^7\Sigma^+$ electronic ground state of the CrBe, CrMg, CrCa, CrSr, CrBa, and CrYb molecules. Points indicate the values for the ground rovibrational level.}
\label{fig:Dip}
\end{figure}

\subsection{Permanent electric dipole moments and static electric dipole polarizabilities}

\begin{figure}
\begin{center}
\includegraphics[width=\columnwidth]{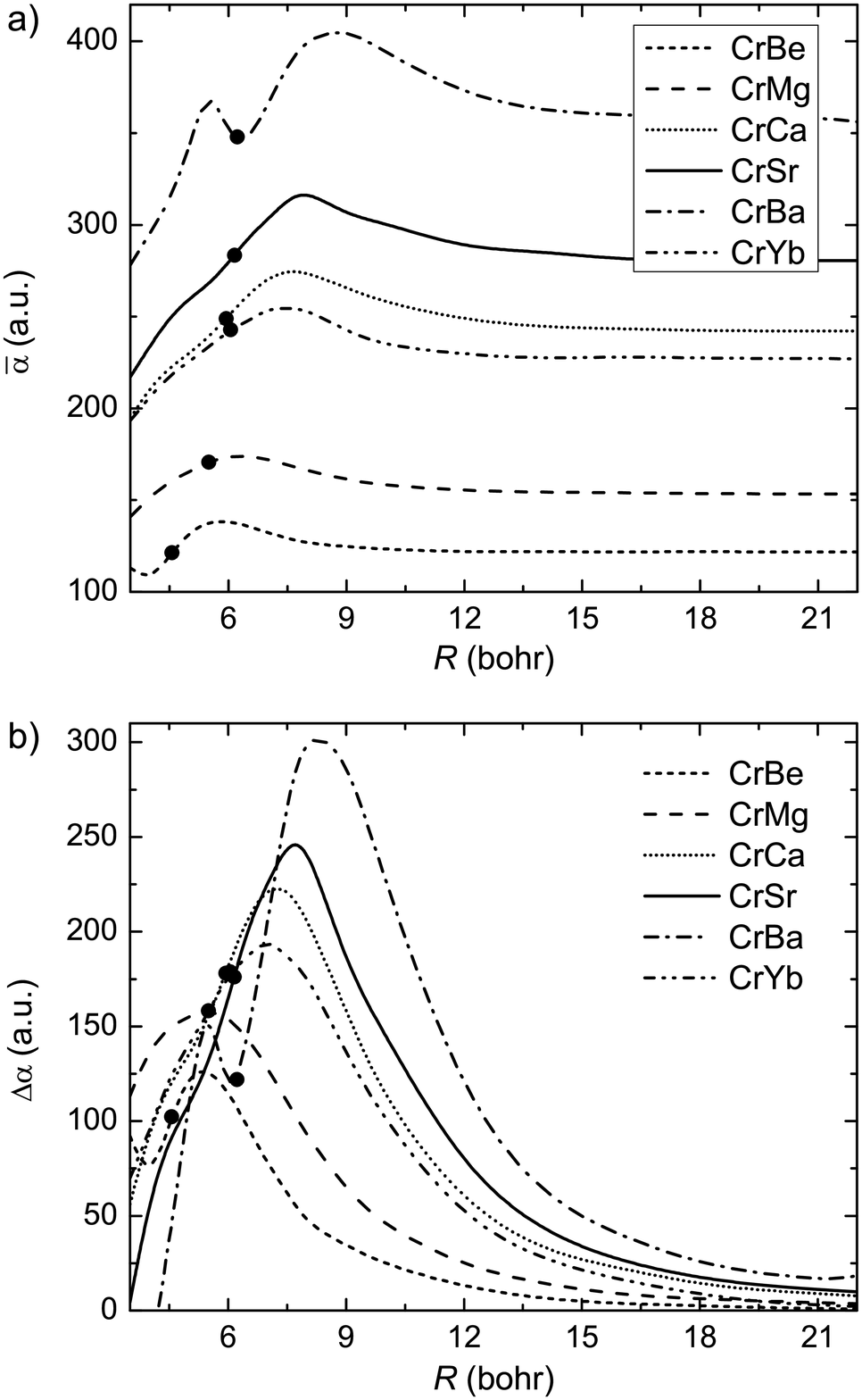}
\end{center}
\caption{The average polarizability (upper panel) and polarizability anisotropy (lower panel) of the $X^7\Sigma^+$ electronic ground state of the CrBe, CrMg, CrCa, CrSr, CrBa, and CrYb molecules. Points indicate the values for the ground rovibrational level.}
\label{fig:Polari}
\end{figure}

Static electric or far-off resonant laser fields can be used to manipulate and control the dynamics of molecules at ultralow temperatures~\cite{QuemenerCR12}. A static electric field couples with an intrinsic molecular electric dipole moment orienting molecules whereas a non-resonant laser field influences the molecular dynamics by coupling with a dipole polarizability anisotropy aligning molecules. 
Both can drastically influence the dynamics and enhance intermolecular interaction, therefore the electric dipole moment and electric dipole polarizability are important properties of ultracold molecules.

The permanent electric dipole moments of the $X^7\Sigma^+$ electronic ground state of the CrBe, CrMg, CrCa, CrSr, CrBa, and CrYb molecules as functions of the interatomic distance $R$ are presented in Fig.~\ref{fig:Dip} and the values for the ground rovibrational level are reported in Table~\ref{tab:Spectro}.

We have found that the CrBa molecule has the largest electric dipole moment in the rovibrational ground state, 2.67$\,$D, only slightly smaller than the CrRb molecule with 2.9$\,$D~\cite{SadeghpourPRA10}. However, the CrSr and CrYb molecules have also significant dipole moments, 1.48$\,$D and 1.19$\,$D, respectively. Since cooling techniques for the Sr and Yb atoms are much further established, the CrSr and CrYb molecules should be considered in the first place as candidates for ultracold molecules with both large magnetic and electric dipole moments. 
The electric dipole moments of the CrSr and CrYb molecules have the values two times larger than the KRb molecule, 0.6$\,$D~\cite{AymarJCP05}, and similar as the RbCs molecule, 1.2$\,$D~\cite{AymarJCP05}, or RbSr molecule, 1.36$\,$D~\cite{BruePRL12}.

There are two independent components of the polarizability tensor for molecules in the $\Sigma$ electronic state, i.e.~the parallel component, $\alpha_\parallel\equiv\alpha_{zz}$, and perpendicular one, $\alpha_\perp\equiv\alpha_{xx}=\alpha_{yy}$. Equivalently, the polarizability anisotropy, $\Delta\alpha=\alpha_\parallel-\alpha_\perp$, and the average polarizability, $\bar{\alpha}=(\alpha_\parallel+2\alpha_\perp)/3$, can be considered.

The average polarizability and the polarizability anisotropy of the $X^7\Sigma^+$ electronic ground state of the CrBe, CrMg, CrCa, CrSr, CrBa, and CrYb molecules are presented in Fig.~\ref{fig:Polari} and the values 
for the ground rovibrational level
are reported in Table~\ref{tab:Spectro}. The polarizabilities show an overall smooth behavior and tend smoothly to their asymptotic atomic values. The interaction-induced variation of the polarizability is clearly visible while changing the internuclear distance $R$.

The polarizability anisotropy, $\Delta\alpha$, is the quantity responsible for the strength of the alignment and the influence of the non-resonant field on the rovibrational dynamics~\cite{GonzalezPRA12,TomzaMP13}.
The larger the average polarizability, $\bar{\alpha}$,
the easier it is to trap molecules e.g.~in an optical lattice.
The CrSr and CrYb molecules have the largest values of the polarizability anisotropy among investigated molecules, 176.1$\,$a.u.~and 178.9$\,$a.u., respectively, in the ground rovibrational state. Therefore, the alignment and control of their dynamics with the non-resonant field should be the easiest and require the lowest field intensity.

In the present work, we have calculated static polarizabilities which describe the interaction of molecules with far non-resonant field, e.g.~from 10$\,\mu$m carbon dioxide laser. When the shorter-wavelength field is applied the dynamic polarizabilities have to be used, which usually are larger but of the same order of magnitude as the static ones.  
Once the wavelength of laser used to control molecules in experiment will be known, the dynamic polarizabilities can be calculated, e.g. from linear response theory~\cite{OlsenJCP85}.

\subsection{Accuracy analysis}

The discussion of the accuracy of the \textit{ab initio} electronic structure calculations requires addressing the following issues:
\begin{itemize}
\item the capability of the computational method to reproduce completely the correlation energy,
\item the completeness of the basis functions set,
\item the relativistic effects.
\end{itemize}

The CCSD(T) method is the gold standard of quantum chemistry and a good compromise between the accuracy and the computational cost~\cite{MusialRMP07}. It reproduces molecular properties such as equilibrium geometries and dissociation energies with the chemical accuracy~\cite{BakJCP01}.  
We have used the spin-restricted RCCSD(T) method in contrast to the existing spin-unrestricted UCCSD(T) method~\cite{KnowlesJCP99} because the spin unrestricted version can potentially lead to the spin contamination for high-spin system such as molecules containing a chromium atom. However, the difference in the interaction energy obtained with two methods is insignificant (less than 2$\,\%$ in the present case). 

Previous calculations for the ground state molecules containing closed-shell atoms reveal that the CCSD(T) method reproduces the potential well depths with an error of a few percent comparing to experimental results. For example, an error for Mg${}_2$ is 0.5$\,\%$~\cite{AmitayPRL11}, for Ca${}_2$ is 1.5$\,\%$~\cite{BusseryPRA03}, and for Sr${}_2$ is 3.8$\,\%$~\cite{SkomorowskiJCP12}. For the two-valence-electron Rb${}_2$ molecule even calculation at the CCSD level gives an error of only 2.7$\,\%$~\cite{TomzaPRA12}. 
However, the chromium atom has six valence electrons in the open shell and we have found that the inclusion of the noniterative triple excitations in the CCSD(T) method accounts for about $30\,\%$ of the interaction energy in the chromium--closed-shell-atom molecules.
The inclusion of full triple or higher excitations in the coupled cluster calculations with high quality basis set for such a large system is computationally unfeasible. Therefore, to estimate the importance of the higher excitations we performed RCCSD(T) and RCCSDT calculation in small aug-cc-pVDZ-DK basis sets for the CrBe, CrMg, and CrCa molecules and we have found that the inclusion of the full triple excitations increases the interaction energy by $7\,\%$, on the average. The lack of the higher excitations should be less important and we estimate the uncertainty of the interaction energy due to the incompleteness of the correlation energy is of the order of $10\,\%$.

The Quintuple-$\zeta$ quality basis sets augmented by the midbond functions used in the present calculations are very extensive computational basis sets that should provide results very close to the complete basis set limit~\cite{SkomorowskiJCP11a}. To evaluate the completeness of them we calculated potential energy curves using the series of the aug-pVnZ-DK basis sets with n=T,Q,5, with and without bond functions. Based on these results we estimate the uncertainty of the interaction energy due to the incompleteness of the basis sets is smaller than $2\,\%$.

The calculation  of the atomic electric dipole polarizability is another check for the quality of the used atomic basis stets and completeness of the method.
The polarizability of the chromium atom from the present calculations is 86.7$\,$a.u., whereas the polarizabilities of the beryllium, magnesium, calcium, strontium, barium and ytterbium atoms are 37.87, 71.7, 158.6, 199.0, 275.5, and $143.9\,$a.u., respectively. These values are in a good agreement with the most sophisticated calculations by Porsev and Derevianko~\cite{DereviankoJETP06}: 37.76, 71.26 , 157.1, 197.2, and 273.5 a.u.~for Be, Mg, Ca, Sr, Ba, respectively, and with the value $143\,$a.u.~for the Yb atom recommended by Zhang and Dalgarno~\cite{ZhangJPCA07}. The polarizability of Cr is in agreement with value 85.0$\,$a.u.~obtained by Pavlovic et al.~\cite{SadeghpourPRA04}.

To evaluate the importance of the relativistic effects on the properties of the considered molecules we additionally calculated potential energy curves with the standard nonrelativistic Hamiltonian and compared them with the ones obtained using the relativistic Douglas-Kroll-Hess Hamiltonian. The well depths are underestimated, on the average by 8$\,\%$, and the equilibrium lengths are longer when the relativistic effects are not accounted for. This is not surprising since the relativistic contribution to the bonding for the transition metal atoms cannot be neglected even for comparatively light chromium atom~\cite{Pelissier86}.
The relativistic effects in the CrBe, CrMg and CrCa molecules were accounted for with the Douglas-Kroll-Hess Hamiltonian whereas for the CrSr, CrBa and CrYb molecules by using energy-consistent pseudopotentials. Therefore, to check the performance of the calculations with ECP  we compared the potential well depths of the CrBe, CrMg and CrCa molecules obtained with the Douglas-Kroll-Hess Hamiltonian with the ones obtained using energy-consistent pseudopotentials. The difference between results obtained with these two methods is of the order of 2$\,\%$, that is much smaller than the relativistic contribution and confirms the validity of the employed approach.  

Based on the above analysis, we estimate that the total uncertainty of the calculated potential energy curves and electronic properties is of the order of $10\,\%$ and the lack of the exact treatment of the triple and higher excitations in the employed CCSD(T) method is a preliminary limiting factor. 

The accuracy of the calculated $C_6$ coefficients is directly related to the accuracy of the input dynamic polarizabilities at an imaginary frequency. For the Be, Mg, Ca, Sr, and Ba atoms they were taken from Derevianko~et al.~\cite{DerevienkoADNDT10} with the accuracy estimated by these authors at the level of 1$\,\%$.
The accuracy of the polarizability of the Yb atom is a few percent. Therefore, the uncertainty of the polarizability of the Cr atom, which is of the order of 5$\,\%$, is a limiting factor for the accuracy of the $C_6$ coefficients.
We estimate that the uncertainty of the computed $C_6$ coefficnets is $5\%$. 
The agreement between the raw \textit{ab initio} data and the asymptotic expansion, Eq.~\ref{eq:E6}, is of the order of 1-3\% at $R\approx30\,$bohr for all investigated molecules.

\subsection{Characteristic energy and length scales}

\begin{figure}
\begin{center}
\includegraphics[width=\columnwidth]{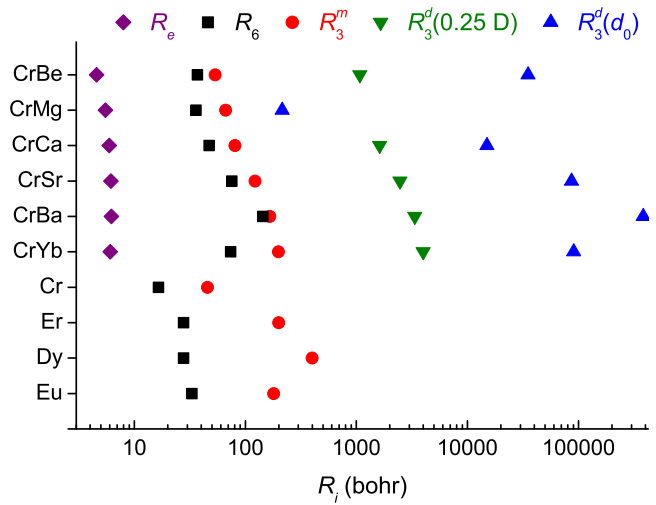}
\end{center}
\caption{(Color online) Characteristic length scales $R_e$, $R_6$, $R_3^m$, and $R_3^d$ for the chromium--closed-shell-atom molecules in the rovibrational ground state and for highly magnetic atoms (electric dipole length, $R_3^d$, for the maximal possible electric dipole moment and for 0.25$\,$D).}
\label{fig:Ri}
\end{figure}

The investigated molecules have both significant magnetic and electric dipole moments. Therefore, to get a good understanding of their collisional properties at ultralow temperatures and the interplay between the electric dipole-dipole, magnetic dipole-dipole, and long-range dispersion interactions 
it is important to understand the various length and energy scales associated with them. 
One can define a characteristic length scale, $R_i$, of the given type of interaction by equating the kinetic energy, ${\hbar^2}/{\mu R_i^2}$, to the interaction potential, $V_i(R_i)$~\cite{JuliennePCCP11}.
The characteristic length scales allow to estimate at what distance a given type of interaction starts to affect the dynamics of  colliding ultracold molecules and to compare the possible influence of different types of interactions on the collisional properties. 
For the electric dipole-dipole, $d_e^2(1-3\cos\theta)/R^3$, magnetic dipole-dipole, $\alpha^2d_m^2(1-3\cos\theta)/R^3$, and Van der Waals, $-\tilde{C}_6/R^6$, interactions the characteristic electric dipole, $R^d_3$, magnetic dipole, $R^m_3$, and van der Waals, $R_6$, lengths are given by:
\begin{eqnarray}
R^d_3&=&\frac{2\mu d_e^2(F)}{\hbar^2}\,,\\
R^m_3&=&\frac{2\mu\alpha^2d_m^2}{\hbar^2}\,,\\
R_6&=& \left(\frac{2 \mu \tilde{C}_6}{\hbar^2}\right)^{1/4}\,,
\end{eqnarray}
where $\mu=m_{\mathrm{CrX}}/2$ is the reduced mass of the pair of molecules, each with mass $m_{\mathrm{CrX}}$, $d_e(F)$ is the induced electric dipole moment at electric field $F$, $d_m=6\mu_B$ is the magnetic dipole moment ($\mu_B$ is Bohr magneton), and $\tilde{C}_6$ is the van der Waals dispersion coefficient for the intermolecular interaction. The $\tilde{C}_6$ coefficients for the interaction between chromium--closed-shell-atom molecules were obtained using simple model:
\begin{equation}
\tilde{C}_6\approx \frac{3}{4}U\bar{\alpha}_0^2+\frac{d_0^4}{6B_0}
\end{equation}
where the first term is the electronic contribution estimated with Uns\"old approximation~\cite{Unsold27} and the second, much larger term, is the contribution from the rotational states calculated assuming molecules in the rovibrational ground state, $v=0,J=0$.  
$U$ is the mean excitation energy, $\bar{\alpha}_0=(\alpha^\parallel_0+2\alpha^\perp_0)/3$ is the mean dipole polarizability, $d_0$ is the electric dipole moment and $B_0$ is the rotational constant of the molecule in the rovibrational ground state.
The computed $\tilde{C}_6$ coefficients are reported in Table~\ref{tab:Spectro}. 

Figure~\ref{fig:Ri} presents the characteristic length scales for the chromium--closed-shell-atom molecules in the rovibrational ground state. The chemical bond length, $R_e$, is the shortest distance. The magnetic dipole length for all species is larger than the van der Waals length, and for the heaviest CrSr, CrYb, and CrBa, it exceeds 100$\,$bohr and is two times larger than for the atomic chromium and of the same order as for the erbium atoms. The electric dipole lengths for the maximal possible dipole moments are much larger than the magnetic dipole lengths. However, the electric dipole moment for a molecule in the rovibrational ground state has to be induced by an external electric field that allows to tune the electric dipole lengths in a wide range of values. Finally, an inspection of Fig.~\ref{fig:Ri} reveals that the intermolecular magnetic dipole-dipole interaction should affect the properties of an ultracold gas of heavy molecules containing chromium atom to a larger extent than it was observed for the ultracold gas of atomic chromium and a competition between magnetic and electric dipolar interactions should be an interesting problem in ultracold many-body physics. 

The stability of an ultracold molecular gas against reactive collisions is an important issue. Since the low-spin Cr$_2$ molecule has a very large binding energy~\cite{SadeghpourPRA04}, much larger than the binding energy of the chromium--closed-shell-atom molecules, there always exists the reactive channel for the collision of two chromium--closed-shell-atom molecules, 
\begin{equation}
2\,\mathrm{CrX}(^7\Sigma^+) \to \mathrm{Cr}_2(^{2S+1}\Sigma_g^+) + \mathrm{X}_2(^1\Sigma_g^+),
\end{equation}
yielding to the chromium molecule in the low-spin state. However, the channel leading to the high-spin Cr$_2$ molecule is closed and one can try to suppress the reactive collisions by applying the magnetic field to restrict molecular dynamics to the maximally spin-stretched electronic potential surface.  
On the other hand, the reactive collisions can be suppressed by applying static electric field to control the long-range dipolar interaction and by
confining molecules in an optical latice to reduce the dimensionality~\cite{QuemenerCR12}.

\section{Summary and conclusions}
\label{sec:summary}

In the present work we have investigated the \textit{ab initio} properties of the chromium--alkaline-earth-metal-atom and chromium--yterbium molecules. Potential energy curves, permanent electric dipole moments, and static electric dipole polarizabilities for the molecules in the $X^7\Sigma^+$ electronic ground state were obtained with the spin restricted open-shell coupled cluster method restricted to single, double, and noniterative triple excitations, RCCSD(T), in the Born-Oppenheimer approximation. The scalar relativistic effects within Douglas-Kroll-Hess Hamiltonian or energy-consistent pseudopotentials were included. The properties of the molecules in the rovibrational ground state were analyzed. The leading long-range coefficients describing the dispersion interaction between the atoms at large interatomic distances, $C_6$, were also computed.

We have found that CrSr and CrYb are the most promising candidates for the ultracold chromium--closed-shell-atom molecules possessing both relatively large electric and large magnetic dipole moments. This makes them potentially interesting candidates for ultracold collisional studies of dipolar molecules in the combined electric and magnetic fields when the magnetic dipole-dipole interaction can compete with the electric dipole-dipole interaction. 
An inspection of the characteristic interaction length scales reveals that the magnetic dipole-dipole interaction for the CrSr and CrYb molecules is of the same order as for the highly magnetic erbium atoms, larger than for the chromium atoms due to larger reduced masses.
The strength of the electric dipole-dipole interaction is controllable as electric dipole moments have to be induced by an external electric field. 
At the same time the large polarizability anisotropy of these molecules allows for the non-resonant light control.

The formation of the proposed molecules will be the subject of a future investigation. Nevertheless, in a similar fashion to the proposals by \.Zuchowski et al.~\cite{ZuchowskiPRL10} and Brue and Hutson~\cite{BruePRL12}, the magnetoassociation using the interaction-induced variation of the
hyperfine coupling constants can be considered in the case of the fermionic $^{53}$Cr atom (provided the widths of the Feshbach resonances are sufficiently broad). On the other hand, a photoassociation near the intercombination line transition of the atomic strontium or ytterbium with the subsequent stabilization into the deeply bound vibrational level of the electronic ground state, similar as predicted for SrYb~\cite{TomzaPCCP11} or Sr$_2$~\cite{SkomorowskiPRA12}, can be proposed. To enhance molecule formation, STIRAP with atoms in a Mott insulator state produced by loading the BEC into an optical lattice~\cite{JakschPRL02} or non-resonant field control~\cite{GonzalezPRA12} can be employed. 

The present paper draws attention to the highly magnetic polar molecules formed from highly magnetic atom and closed-shell atom and paves the way towards a more elaborate study of the magneto or photoassociation and application
of these polar and magnetic molecules in ultracold many-body physics studies.

\acknowledgments
We would like to thank Robert Moszynski, Piotr \.Zuchowski and Jedrzej Kaniewski for many useful discussions.
Financial support from the Polish Ministry of Science and Higher
Education through the project N N204 215539
and from the Foundation for Polish Science within
MPD Programme co-financed by the EU European
Regional Development Fund is gratefully acknowledged.

\end{document}